\documentclass[9pt,twocolumn,twoside]{osajnl}

\journal{ol} 

\setboolean{shortarticle}{true} 

\title{Electro-mechanical control of an on-chip optical beam splitter containing an embedded quantum emitter}

\author[1,*]{Z.K. Bishop}
\author[1]{A.P. Foster}
\author[1]{B. Royall}
\author[1]{C. Bentham}
\author[2]{E. Clarke}
\author[1]{M.S. Skolnick}
\author[1]{L.R.~Wilson}

\affil[1]{Department of Physics and Astronomy, University of Sheffield, Sheffield S3 7RH, United Kingdom}
\affil[2]{Department of Electronic and Electrical Engineering, University of Sheffield, Sheffield S1 3JD, United Kingdom}

\affil[*]{Corresponding author: z.k.bishop@sheffield.ac.uk}

\ociscodes{(130.4815) Optical switching devices; (230.4685) Optical microelectromechanical devices; (270.5585) Quantum information and processing.}

\doi{\url{https://doi.org/10.1364/OL.43.002142}}

\begin{abstract}
We demonstrate electro-mechanical control of an on-chip GaAs optical beam splitter containing a quantum dot single photon source. The beam splitter consists of two nanobeam waveguides, which form a directional coupler. The splitting ratio of the directional coupler is controlled by varying the out-of-plane separation of the two waveguides using electromechanical actuation. We reversibly tune the beam splitter between an initial state, with emission into both output arms, and a final state with photons emitted into a single output arm. The device represents a compact and scalable tuning approach for use in III-V semiconductor integrated quantum optical circuits.
\end{abstract}

\setboolean{displaycopyright}{true}

\begin{document}

\maketitle

Micro-opto-electro-mechanical systems (MOEMS) have been widely studied for a variety of applications in semiconductor integrated photonic circuits.  The vast majority of work has been carried out in silicon where on-chip tuning of the optical properties of essential circuit components is possible by displacing them mechanically with the application of an electrostatic field.  Lateral displacement has been used in phase modulators~\cite{Acoleyen2012,Winger2011}, resonance tuning of nanobeam photonic-crystal cavities (PhCCs)~\cite{Deotare2009,Frank2010} and microtoroid resonators~\cite{Baker2016}.  More complicated structures such as comb-drive actuators have also been developed to allow for larger  displacements~\cite{Legtenberg1996,Zhou2003,Chew2010,Chew2010-1,Shi2015}, attractive for optical switching applications~\cite{Bulgan2008,Akihama2011,Akihama2012,Munemasa2013}.  Recently, scalable out-of-plane actuation methods have also been demonstrated at room temperature based on a cantilever geometry~\cite{Han2015}. 

MOEMS based on III-V semiconductors are now emerging for applications in quantum information processing (QIP).  Initial work has focused on tuning of PhCC modes into resonance with quantum emitters in order to enhance their emission.  In-plane~\cite{Ohta2013} as well as double-membrane out-of-plane actuation methods have been reported~\cite{Midolo2011,Midolo2012,Petruzzella2015}.  Beam splitters, realized on chip using directional couplers (DCs), are another key component of integrated linear quantum optical circuits, with post-fabrication control of their optical properties likely required for efficient QIP applications~\cite{Knill2001,Kimble2008}. In this context, electro-mechanical tuning of DCs has so far only been considered theoretically by Liu et al.~\cite{Liu2017}, using a double-membrane actuation approach.

In this Letter we demonstrate the electro-mechanical control of an on-chip beam splitter operating at low temperature probed using single photon emission from an embedded InGaAs quantum dot (QD).  The proposed device is compact, easy-to-fabricate and scalable~\cite{Han2015} with large achievable out-of-plane displacements of over $400$\,nm.  The device structure is versatile and can be adapted to fine tune other on-chip photonic elements.  It represents a significant step towards reconfigurable integrated quantum optical circuits with embedded single photon sources.

\begin{figure}
\centering
\includegraphics[width=1\linewidth]{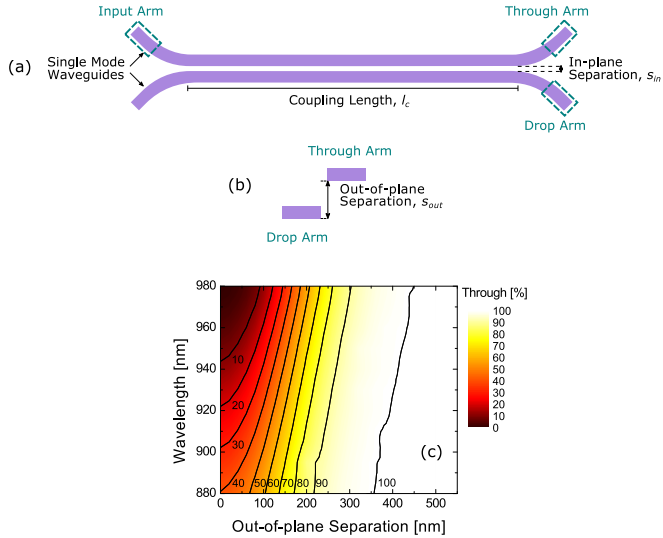}
\caption{\label{Fig01}(a) Top- and (b) side- view schematic diagram of a nanobeam waveguide directional coupler. (c) Results of the modeling of a directional coupler consisting of $160$\,nm thick and $280$\,nm wide waveguides, separated laterally by $40$\,nm in a $7\,\mu$m long coupling region.  The contour plot shows how the fraction of light evanescently coupled from one channel to the other depends on the wavelength of the transmitted light and the out-of-plane separation between the waveguides.}%
\end{figure}

The operating principle of our device is shown schematically in Fig.\,\,\ref{Fig01}.  The DC acts as an optical beam splitter for light entering the input arm, due to evanescent light coupling between the two waveguides in the coupling region. The ratio of the output power in the through and drop arms of the DC is defined as the splitting ratio (SR). The SR depends on the dimensions of the waveguides, the wavelength of the transmitted light, and both in-plane, $s_{in}$, and out-of-plane, $s_{out}$, separations between the waveguides.  Here, we tune the parameter $s_{out}$ in order to control the SR of the beam splitter.

To determine the theoretical change in the SR of the DC as $s_{out}$ is varied, $2$D electromagnetic modeling was undertaken using MIT Photonic-Bands, a freely available eigenmode solver.  The results for a range of wavelengths are shown in Figure\,\ref{Fig01}(c) for a DC consisting of $160$\,nm thick and $280$\,nm wide single (TE) mode waveguides, separated laterally by $s_{in}=40$\,nm in a $7$\,$\mu$m long coupling region ($l_{c}$). It is clear that in a broadband QD emission wavelength range of $880-980$\,nm, the DC can be tuned from an overcoupled state, when more light is coupled to the drop than the through arm, to a decoupled state, when all the light is transmitted to the through arm, as $s_{out}$ is increased to an achievable $400$\,nm.  This demonstrates the potential of the proposed device, which allows the SR to be switched between the commonly required $50$:$50$ and an output into a single arm, $100$:$0$.  Moreover, $3$D finite-difference time-domain electro-magnetic simulations were also performed to confirm that the high evanescent coupling efficiency of $98$\,\% between the two waveguides varies by less than $0.5$\,\% over the accessible range of $s_{out}$.

\begin{figure}[t]
\centering
\includegraphics[width=1\linewidth]{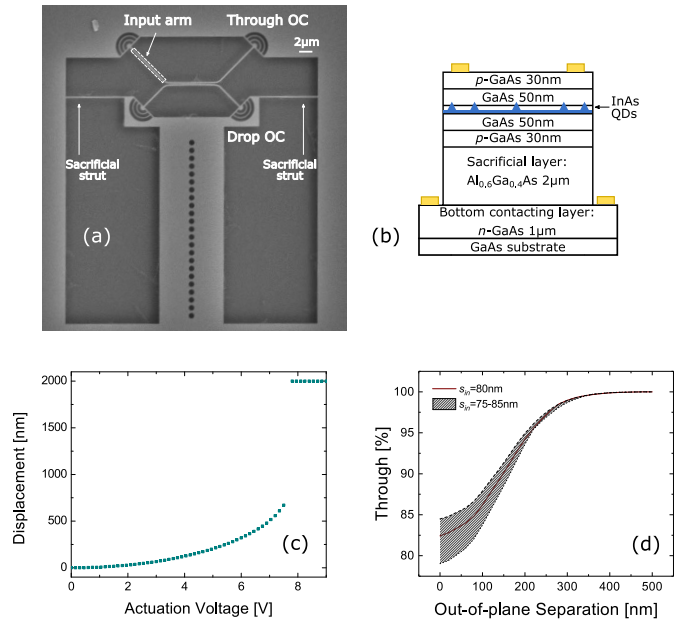}
\caption{\label{Fig0}(a) Top-view scanning electron microscope image of a typical device. The sacrificial struts are removed at low temperature before the opto-electro-mechanical measurements.  (b) Schematic diagram of the wafer structure.  The position of contacts is marked by gold rectangles.  (c) Calculated displacement of the free end of the $35\,\mu$m long and $7.5\,\mu$m wide cantilever as actuation voltage is increased. (d) Modeled overall percentage of light transmitted to the through output-coupler with increasing out-of-plane waveguide separation for the experimental device with $s_{in}=80\pm5$\,nm.}
\end{figure}

Experimentally, we control $s_{out}$ by attaching one of the nanobeam waveguides of a DC to a mechanically compliant cantilever, which is actuated electro-mechanically to induce out-of-plane waveguide separation. Figure\,\ref{Fig0}(a) shows our proof-of-concept device, consisting of a GaAs nanobeam waveguide DC with one waveguide attached to the free end of a $35$\,$\mu$m long and $7.5$\,$\mu$m wide cantilever, and the other fixed rigidly to the bulk of the sample.  Each arm is terminated with a Bragg output-coupler (OC) to enable out-of-plane collection of transmitted photons.  The free end of the cantilever is supported on either side by a $300$\,nm wide sacrificial strut during sample fabrication and transport.  The struts are removed before measurements commence using local laser ablation, releasing the cantilever.

The device was fabricated on a {\it{p-i-p-i-n}} diode, the schematic of which is presented in Fig.\,\ref{Fig0}(b).  The DC and the cantilever were defined within the $160$\,nm thick top {\it{p-i-p}} GaAs membrane using electron-beam lithography followed by an inductively coupled plasma etch.  The intrinsic region of this membrane contained InGaAs self-assembled QDs, used as embedded single photon sources to probe the optical response of the system.  The {\it{n}}-GaAs substrate was electrically isolated from the membrane by a $2\mu$m thick intrinsic Al$_{0.6}$Ga$_{0.4}$As layer, which was removed from underneath the device using an HF etch to create the suspended structure.  The stresses that may occur in the structure due to surface tension when drying the device in air afterwards were minimized using a critical point drying technique.  In this method the rinsing water was purged with liquid CO$_2$ and the sample was brought to the temperature and pressure critical for CO$_2$, allowing to dry the device without surface tension present.  Ni:Au contacts were made to the top {\it{p}}- and the bottom {\it{n}}-GaAs layers in order to allow for electro-mechanical control of the cantilever.  The row of holes in the center of the device [see Fig.\,\ref{Fig0}(a)] allowed for faster under-etching of the cantilever.

Applying an actuation voltage, $V_{act}$, between the cantilever and the substrate results in a capacitive force, which causes the cantilever to deflect towards the substrate.  This introduces a vertical out-of-plane separation between the two arms of the DC.  Figure\,\ref{Fig0}(c) shows the displacement of the free end of the cantilever as $V_{act}$ is increased, calculated using an analytical model which determines the displacement for a given $V_{act}$ by minimizing the total energy of the system (comprising strain and electrostatic energies).  The model assumes that the cantilever's vertical displacement is a quadratic function of position along its length~\cite{Tsuchitani1998}. The theoretical maximum controllable displacement of the cantilever is $1/3$ of the initial distance between the cantilever and the substrate, $s_{0}$~\cite{Midolo2014}.  Once this displacement is reached at the so-called pull-in voltage, $V_{pull}$, the capacitive force becomes greater than the restoring force and the free end of the cantilever collapses onto the substrate.  This introduces surface adhesion forces between the cantilever and the substrate.  If these forces are smaller than the restoring force of the cantilever, the cantilever will be able to lift back up from the substrate at $V_{act}<V_{pull}$ resulting in a hysteresis behavior~\cite{Lee2013,Buchnev2016}.  For our system $V_{pull}$ is calculated to occur at $7.5$\,V, when the discontinuity in the filled squares curve is observed between the displacement of $667$\,nm and $2000$\,nm (corresponding to $s_{pull}=s_{0}/3$ and $s_{0}$).

Our proof-of-concept device had a waveguide width of $280$\,nm and $s_{in}=80\pm5$\,nm.  The latter is $\sim40$\,nm larger than the target value of $40$\,nm and is caused by fabrication inaccuracies, which are more pronounced for smaller separations.  Figure\,\ref{Fig0}(d) shows the theoretical output from the through arm of the DC with these dimensions as $s_{out}$ is increased.  The operation wavelength chosen is $910$\,nm as this is the emission wavelength of the single QD studied experimentally. The initial SR is $\sim83$:$17$, and can be increased to $100$:$0$ for $s_{out}>300$\,nm. The model is in good qualitative agreement with our experimental results, as we show below.

The electro-mechanical behavior of the cantilever was studied experimentally using white light illumination of the sample in a cryostat at $4.2$\,K using a confocal microscope system and an infrared camera.  As $V_{act}$ was increased the free end of the cantilever was observed to collapse onto the substrate at $13$\,V, and then lift back up as $V_{act}$ was decreased to $4$\,V, allowing for multiple measurements to be performed.  This was possible due to two factors.  Firstly, large $s_0$ means a large restoring force of the cantilever, which counters the surface adhesion forces present after the collapse.  Secondly, only a small portion of the free end of the cantilever is actually in contact with the substrate, minimizing the action of these forces.  The larger $V_{act}$, compared to the modeling, required for the actuation of our cantilever could simply be related to the resistance of the contacts, which is not taken into account in the modeling.

\begin{figure}[t]
\centering
\includegraphics[width=1\linewidth]{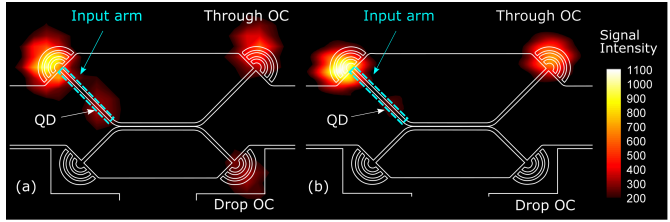}
\caption{\label{Fig2} Filtered PL collection map of the device with an overlaid device contour with (a) $V_{act}=0$\,V, and (b) $V_{act}=12.5$\,V.}
\end{figure}

The device was studied optically using micro-photoluminescence ($\mu$-PL) spectroscopy with spatially resolved excitation and collection in an exchange gas cryostat at $4.2$\,K using a confocal microscope system.  For the measurements of the SR we selected a bright and spectrally isolated QD embedded within the input arm of the fixed waveguide of the DC, emitting at $910.6$\,nm.  The QD was excited from above via the wetting layer using a Ti:Sapphire CW laser emitting at $840$\,nm.  Figure\,\ref{Fig2} shows two $\mu$-PL maps, with the device contour overlaid, obtained by raster scanning the collection across the device while spectrally filtering at the QD wavelength.  Figure\,\ref{Fig2}(a) was obtained from the device with $V_{act}=0$\,V and emission can be seen from both the through and drop OCs.  The $\mu$-PL map in Fig.\,\ref{Fig2}(b) was acquired for the device operated with $V_{act}=12.5$\,V, and emission from the drop OC is observed to be heavily suppressed, while that from the through OC increases as expected.  

\begin{figure}[t]
\centering
\includegraphics[width=1\linewidth]{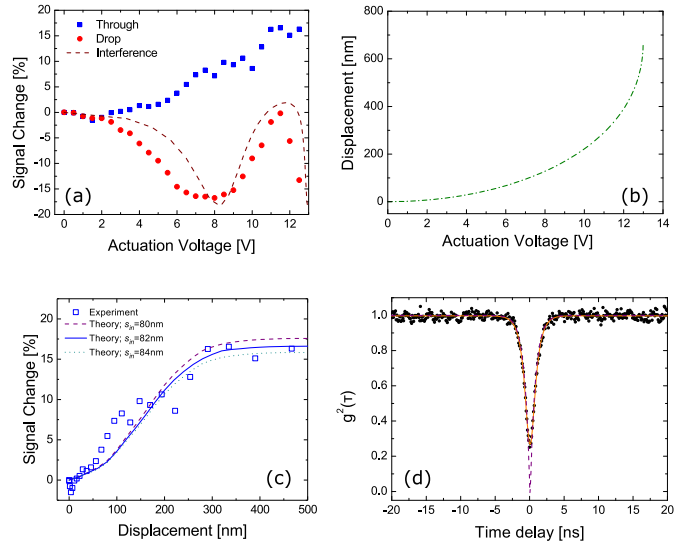}
\caption{\label{Fig3}(a) Measured changes to the QD signal collected from the through and drop OCs independently, as actuation voltage is increased. The signal is normalized to the total signal recorded from both OCs at $V_{act}=0$\,V.  The peak in the signal from the drop OC at about $V_{act}=11$\,V is due to changing optical interference from the moving OC during cantilever actuation (modeled using the transfer-matrix method).  (b) Cantilever displacement as a function of actuation voltage for the measured system, found using Eq.\,(\ref{Eq2}) with $X=2650$.  (c) Experimental results (empty squares) for the through OC from graph (a) as a function of displacement converted from actuation voltage using the relationship in Fig.\,\ref{Fig3}(b).  The other three lines are theoretical curves for $s_{in}$ of $80$, $82$, and $84$\,nm normalized to the initial signal at zero displacement.  (d)~Normalized second-order correlation function obtained by exciting the QD from above and collecting the spectrally filtered PL signal from the input OC. The orange continuous line is a fit to the experimental data (black points), while the dashed purple line represents a fit that takes into account the time response of the measurement system.}
\end{figure}

To characterize the device at increasing out-of-plane waveguide separation, the routed QD emission was measured simultaneously from the through (fixed) and the drop (moving) OCs using two independent collection paths as $V_{act}$ was increased.  The SR for the device at $V_{act}=0$\,V was measured to be $80$:$20$.  The absolute percent change to the measured signal is shown in Fig.\,\ref{Fig3}(a) for the through and drop OCs separately.  The signal is normalized to the total signal collected from both OCs at $V_{act}=0$\,V.  The change in the QD emission collected from the through OC increases monotonically until it saturates at $17$\% for $V_{act}>11$\,V.  The signal at the drop OC decreases initially as expected, but then recovers and peaks at about $V_{act}=11.5$\,V before decreasing again.  This is caused by the downward movement of the drop OC as $V_{act}$ is applied, which results in changes to the optical interference of signal emitted from the OC and that reflected from the substrate as well as collection efficiency changes.  While the optical interference effect can be modeled using the transfer-matrix method, which explains the peaks and troughs in the signal measured from the drop OC (see Fig.\,\ref{Fig3}(a)), the collection efficiency changes are more difficult to estimate.  In addition, without performing fully coupled opto-electro-mechanical simulations it is not possible to deconvolve these effects acting on the drop OC from the evanescent coupling that is of interest.  Hence, we proceed to determine the controlled changes to the splitting ratio based on the through OC only.  A small QD Stark-shift of $0.15$\,nm was also observed as $V_{act}$ was increased from $0$\,V to $12.5$\,V, due to the increasing electric field between the top {\it{p}} doped layer of the membrane and the {\it{n}} doped substrate. Contacting the lower {\it{p}} doped layer of the GaAs membrane would eliminate the Stark shift, and would enable simultaneous electro-mechanical actuation and QD Stark tuning~\cite{Petruzzella2015}.

In order to directly compare the controlled optical properties of the DC with the modeling [shown in Fig.\,\ref{Fig0}(d)], $V_{act}$ was converted to cantilever displacement.  To do so, we balanced the capacitive force between the cantilever and the substrate with the restoring force of the cantilever to obtain:

\begin{equation}
\frac{\epsilon_{0}A}{2(s_{0}-s_{out})^2}V_{act}^2=ks_{out},
\label{Eq1}
\end{equation}
where $s_{0}$ is the initial distance between the cantilever and the substrate (here $2000$\,nm), $s_{out}$ is the displacement of the cantilever, $\epsilon_{0}$ is the permittivity of free space, $A$ is the surface area of the cantilever, and $k$ is a fitting parameter corresponding to an average stiffness of the cantilever.  We thus derived the relationship between $V_{act}$ and the displacement as follows:

\begin{equation}
V_{act}=\frac{(s_{0}-s_{out})\sqrt{s_{out}}}{X},
\label{Eq2}
\end{equation}
where $X=\sqrt{\epsilon_{0}A/2k}$ \,nm\textsuperscript{3/2}V\textsuperscript{-2}.

The parameter $X$ for our measured device was found using $V_{pull}=13$\,V (the observed pull-in voltage) and the theoretical maximum displacement of $s_{pull}=667$\,nm.  Equation\,(\ref{Eq2}) with $X=2650$ was then used to convert $V_{act}$ to cantilever displacement and the resulting curve is shown in Fig.\,\ref{Fig3}(b).

This conversion allows us to directly compare the experimentally determined variation in the QD emission routed to the through OC with that calculated using the eigenmode solver [the results of which are presented in Fig.\,\ref{Fig0}(d)].  The experimental curve is in the best agreement with the theoretical curve for $s_{in}=82$\,nm, presented in Fig.\,\ref{Fig3}(c) as a function of cantilever displacement.  The two curves demonstrate monotonic increase of the QD signal until saturation at $17$\,\% for displacements of over $300$\,nm.  The achieved displacement of the cantilever was found to be over $400$\,nm before the pull-in occurs.

We verified the single-photon nature of the emission from this QD by performing an on-chip Hanbury Brown and Twiss experiment, which consists of cross-correlating the photons at the QD wavelength collected by two separate paths from the input OC.  The results are shown in Fig.\,\ref{Fig3}(d), with normalized $g^{(2)}(0)=0.25\pm0.02$.  By deconvolving the experimental data with the temporal response of our detection system (Gaussian, full-width-at-half-maximum of $874\pm4$\,ps) we obtain $g^{(2)}(0)=0\pm0.01$, which indicates that the source is strongly antibunched.

The electro-mechanical system presented here can not only be used to control the SR of an on-chip beam splitter but also to fine tune other integrated photonic devices.  Greater versatility and scalability of the system can be achieved through some improvements to the sample design~\cite{Han2015}, which could enable it to operate as an optical router with an expected switching rate of the order of $0.5$\,MHz.  Further optimization of the dimensions of the DC can overcome the difficulties in achieving small enough in-plane separations needed in the reported device for larger tuning range covering the commonly required $50$:$50$ splitting.  Increasing the coupling length of the DC or decreasing the cross section of the waveguides are examples of promising approaches worth investigating.  Operating the beam splitter at longer telecoms wavelengths is another solution to achieving 50:50 splitting with larger and easier to achieve in-plane waveguide separations.  Fabricating the structure on a {\it{p-i-n-i-n}} diode and depositing a third contact on the middle {\it{n}}-layer would enable tuning of the QD emission wavelength using the quantum-confined Stark effect~\cite{Bentham2015} at the same time as controlling the beam splitter electro-mechanically.  The operating actuation voltage could also be decreased by using a thinner AlGaAs sacrificial layer, reducing the initial distance between the two electrodes.  Alternatively, if the device footprint is of importance, the AlGaAs thickness could be decreased in order to achieve the same electro-mechanical performance for a shorter cantilever.  However, a thinner AlGaAs layer may affect the observed recovery of operation after pull-in.

In conclusion, we have demonstrated electro-mechanical control of an on-chip beam splitter operating at low temperature using out-of-plane actuation, with large achievable displacements of over $400$\,nm.  An embedded quantum emitter was used to probe the optical response of the system.  The splitting ratio of our on-chip optical beam splitter was tuned from an initial $\sim80$:$20$ at $V_{act}=0$\,V (zero displacement) up to $\sim100$:$0$ at $V_{act}=11$\,V ($300$\,nm displacement).  The proposed device operates as a fine tuning element and paves the way toward increased control of on-chip single photon devices using compact, easy-to-fabricate and scalable structures for use in III-V semiconductor integrated quantum optical circuits.

This work was funded by EPSRC Grants No. EP/J007544/1 and EP/N031776/1.  The data from this study is available at 10.15131/shef.data.6022892.

\bibliography{Ref}

\bibliographyfullrefs{Ref}

\end{document}